\documentstyle[aps,prl,preprint]{revtex}

\begin{document}
\author{
Dongsheng Du${}^{1,2}$~~~ Hongying Jin${}^2$~~~  Yadong Yang${}^{1,2,3}$
\footnote{Email: duds@bepc3.ihep.ac.cn, yangyd@bepc3.ihep.ac.cn}\\
{\small\sl ${}^{1}$ CCAST (World Laboratory), P.O.Box 8730, Beijing 
100080, China}\\
{\small\sl ${}^{2}$ Institute of High Energy Physics, Academia Sinica,
P.O.Box 918(4), Beijing 100039, China\thanks{Mailing address} }\\
{\small\sl ${}^{3}$ Physics Department of Henan Normal University, Xingxiang,
Henan, 453002, China }
}
\date{}
\title{
{\large\sf
\rightline{BIHEP-Th/97-007}
}
\vspace{3cm}
\bigskip
\bigskip
{\LARGE\sf Probe New Physics in Leptonic $ B_c$  Decays at CERN LHC}
  }
\maketitle

\begin{abstract}
\noindent
With respect to large samples of $B_c$ mesons expected to be produced at
CERN Large Hadron Collider(LHC) and the large branching ratios
$Br(B_{c}\rightarrow \tau\bar\nu)$ and $Br(B_{c}\rightarrow \mu\bar\nu),$ 
we suggest that $B_c$ purely leptonic decays
 could offer an unique probe of the standard model and  its extensions 
 such as two Higgs doublet models and models involving supersymmetry, and
 also  the structure of  charged weak currents.

\end{abstract}
\vspace{1.5cm}
{\bf PACS  numbers 13.20.He 12.15.-y 12.60.Fr} 

\newpage

Being consisted of two different heavy flavors, $B_c$ meson is believed 
to offer
an unique probe of both strong and weak interactions. The physics of  $B_c$
meson has stimulated much recent works on their properties,
weak decays and production cross section at high energy colliders.
Although $B_c$ meson is hard to be produced, the number of $B_c$ meson
produced at LHC  is estimated[1] to be $2.1\times 10^8 $ (for $100fb^{-1} $
integrated luminosity with cuts of
$P_T (B_c )> 20 {\rm GeV}, \mid y(B_c )\mid <2.5 $ ).
Once produced, an excited $b\bar c$ meson will cascade down through lower energy
$b\bar c$ states via hadronic or electromagnetic transitions to
the pseudoscalar ground state $B_c$ with mass lying below the ${\bar B}D$
threshold, which decays weakly. The weak decays 
$B_{c}\rightarrow \tau\bar\nu$, $\mu \bar\nu$
 are of particular
interest because of at least two reasons:

{\bf $\bullet$} The compact size of $B_c$ meson enhanced the importance of
annihilation decays, $i.e$, the decay constant $f_{B_c }$ is much larger
than that of other heavy mesons.  Furthermore, compared with
$\Gamma( B_u^{-} \rightarrow l^- {\bar \nu_{l}}) $,
$\Gamma( B_c^{-} \rightarrow l^- {\bar \nu_{l}})$ is enhanced by a large factor
\begin{equation}
\frac{\Gamma( B_c^{-} \rightarrow l^- {\bar \nu_{l}}) }
{\Gamma( B_u^{-} \rightarrow l^- {\bar \nu_{l}})}=\left( \frac{f_{B_c}}{f_{B_u}}
\right)^2
\left|\frac{V_{cb}}{V_{ub}}\right|^2 \frac{M_{\rm B_c}}{M_B}\approx 10^3\sim 10^4 .
\end{equation}
Recently,  the development of Heavy Quark Effective Theory(HQET)[2]
 has  placed the extraction of $V_{cb}$ on a solid theoretical
ground,
 on which the extraction of $V_{cb}$ can be treated in a nearly
 model-independent way[3], however, it is  not the case 
for the extraction of $V_{ub}$. \par
{\bf $\bullet\bullet$ } Purely leptonic decays of $B_c$ are sensitive
to new physics beyond the SM at tree level.

The leptonic decay of $B_c$ in the SM proceeds through a virtual
W as in Fig.1. All QCD effects, both perturbative and nonperturbative,
enter into the decay rate though the decay constant $f_{B_c}$, defined 
by the matrix element 
\begin{equation}
\langle 0| {\bar c}\gamma_{\mu}\gamma_5 b|B_{c}(P) \rangle=-if_{B_c}P_{\mu}.
\end{equation}

In terms of the decay constant, $ B_c$ leptonic decay width  in the SM is 
\begin{equation}
\Gamma( B_c^{-} \rightarrow l^- {\bar \nu_{l}})
=\frac{G_F^2}{8\pi}|V_{cb}|^2 M_{B_c} f_{B_c}^2 m_l^2 \left( 1-
\frac{ m_l^2}{ M_{B_c}^2}\right)^2 .
\end{equation}

The formula (2) provides a nonperturbative definition of the decay constant
$f_{B_c}$, so that it can be calculated using lattice QCD simulation.
One of the difficulties with such a calculation is that it requires a lattice
with large volume and fine lattice spacing, since the strong interactions
must be accurately simulated over many distance scales. 

Braaten and
Fleming[4] have suggested an elegant way  for calculating $f_{B_c}$
using nonrelativistic QCD(NRQCD)[5] which is a rigorous theory for heavy
quarknium annihilation.
Using the factorization formalism  developed in NRQCD, 
one can separate short-distance
and long-distance effects, and then  the short-distance effects can be
calculated  analytically using perturbative theory in $\alpha_s$. On the
other hand, the nonperturbative effects can be studied symmetrically order by
order in terms of  the expansions in NRQCD. 
The remaining  matrix can be worked
out using a much coarser lattice which provides enormous saving in computer
resources.

In a word, benefiting from the results from both HQET and NRQCD,
the $B_c$ leptonic decay width could be inevitably predicted  accurately.
So we suggest that it would offer an unique theoretical
clean testing ground for the SM and its extensions
such as two Higgs doublet models and models involving supersymmetry, and
also its  $(V-A)$ structure of charged weak currents.\par

At first, we will examine charged Higgs effects in
$ B_c^{-} \rightarrow l^- {\bar \nu_{l}}$ ($l=\mu,~\tau$). 
We take the so-called Model II
of two Higgs doublet models[6] in 
which one Higgs doublet couples to down-type 
quarks and charged leptons and the other to up-type quarks. The minimal 
supersymmetric standard model belongs to this class, so the following  
results also apply to this model.

The Yukawa interaction of the charged physical scalars $H^{\pm}$
 with fermions
is determined by $tan\beta$ (the ratio of the vacuum expectation values of
the two  Higgs doublets ), by the  fermions masses and CKM matrix. The terms
in the Lagrangian relevant for $ B_c^{-} \rightarrow l^- {\bar \nu_{l}}$ 
are 
\begin{equation}
{\cal L}_{eff}=-V_{cb}\frac{4G_F}{\sqrt{2}}
\left[ 
(\bar c \gamma_{\mu}P_L b)(\bar l \gamma^{\mu}P_L \nu_{l})
-R(\bar c P_R b)(\bar l P_R \nu_{l}) 
\right] ,
\end{equation}
where 
\begin{equation}
R=r^2 m_l m_b ,~~~~~~~~r=\frac{tan\beta}{M_{H^\pm }} ,
\end{equation}
and $P_{L,R}=\frac{1}{2}(1\mp \gamma_5 )$.
We have neglected a term proportional to $m_c$ as in the literature[8],
because it is suppressed by the mass  ratio $m_c /m_b$ and can not be
enhanced by the possible large factor $tan^2 \beta$.

One obtains 
\begin{equation}
\Gamma( B_c^{-} \rightarrow l^- {\bar \nu_{l}})=
\frac{G_F^2}{8\pi}|V_{cb}|^2 M_{B_c} f_{B_c}^2 m_l^2 \left( 1-
\frac{ m_l^2}{ M_{B_c}^2}\right)^2  \times
\left( 1-tan^{2}\beta \frac{M_{B_c}^2}{ M_{H^\pm }^2}\right)^2 ,
\end{equation}
where we have used the relation 
\begin{equation}
\langle 0| {\bar c}\gamma_5 b|B_{c}(P) \rangle=-if_{B_c}
\frac{M_{B_c}^2}{m_b +m_c }\approx -if_{B_c}M_{B_c}.
\end{equation}

It was noted by Hou[7] that the $H^{\pm}$ boson simply modified the
SM prediction  of $Br(B_u^- \rightarrow \l^- \nu_{\l})$ by a $m_l$-independent factor.
Certainly, it is also true in the present case.  Actually, the charged Higgs
effects in $B_{u,d}$ simeleptonic decays have been discussed in the
literature[8].
Recently, using a data sample of 1,475,000 $Z q\bar q (\gamma)$ 
events, L3 collaboration[9]  has studied the purely leptonic decays 
of heavy flavor mesons, $D_s^- \rightarrow \tau^- \nu_{\tau}$, 
$B_u^- \rightarrow \tau^- \nu_{\tau}$. No signal of 
$B_u^- \rightarrow \tau^- \nu_{\tau}$   
is observed in the data, yielding the upper limit 
\begin{equation}
Br(B_u^- \rightarrow \tau^- \nu_{\tau})< 5.7\times 10^{-4}. 
\end{equation}
Assuming $f_B =190 {\rm MeV} $ and using $V_{ub}=0.0033\pm0.0008$[10], they got
the following constraint 
\begin{equation}
r=\frac{tan\beta}{M_{H^\pm}}<0.38 {\rm GeV}^{-1}, ~~~~at~~ 90\%~~CL, 
\end{equation}
which approaches the best limits on $tan\beta$
and $M_{H^\pm}$ from the proton stability experiment[11] and from measurements
of $b\rightarrow s\gamma$ transition[12].
It needs to be noted that there exists large uncertainties in $V_{ub}$
and it is well known that the extraction of $V_{ub}$ is very difficult because
of both theoretical and experimental reasons.
\par
In Fig.2, we  display the charged Higgs effects in
$\Gamma(B_c^- \rightarrow \tau^- \nu_{\tau})$ in the range $0<r<0.38{\rm GeV}^{-1}$.
It is seen that $\Gamma(B_c^- \rightarrow \tau^- \nu_{\tau})$  is very sensitive
to $r$.
Using  $\tau_{B_c}=0.52$[13], it is estimated in the SM that
$Br(B_c^- \rightarrow \tau^- \nu_{\tau})=3\%$ and
$Br(B_c^- \rightarrow \mu^- \nu_{\mu})=10^{-4}$. Given $10^{8}~~B_c $ mesons
produced at LHC, there will be about $10^6$ events of
$ B_c^- \rightarrow \tau^- \nu_{\tau}$ and $10^4$ events of
$ B_c^- \rightarrow \mu^- \nu_{\mu}$.
It is foreseen that LHC will run at
${\cal L} \approx 1\times 10^{33}cm^{-2}s^{-1}$ during the first year and the
luminosity will increase later on and the expected number of $B_c$ mesons
can reach $10^{10}$ per year[14]. Even $0.1\% $ of the events are useful,
it is still possible to measure the channels $ B_c^- \rightarrow \tau^- \nu_{\tau}$
 accurately.\par
As argued  above, purely leptonic $B_c$ decays is a good window to test
the SM and its extensions. In what following, we will examine the effect of
possible admixture of  $(V+A)$  current $g_R (\bar{c}_R\gamma_{\mu}b_R ) $
to the standard $(V-A)$ current $g_L (\bar{c}_L\gamma_{\mu}b_L ) $.
The possibility of a presence of such non-$V-A$ structures has been
 most extensively
explored for the muon decay, while for the $b$ decays so far  only the
maximal case of a $(V+A)\times (V-A)$ structure of the four fermions interaction
 is excluded experimentally[15] and another extreme of a purely vector
 $b\rightarrow c$ current is clearly excluded by the very fact of non-zero
 amplitude of the decay $B\rightarrow D^* l\nu$ at zero recoil. A small value of
 $g_R /g_L$ is not ruled out as discussed in ref.[16] and 
 can be sought for  as one possible sign of new physics.
 
 The structure of the four-fermions interaction 
for our concern can be written as
 \begin{equation}
{\cal L}_{eff}=-V_{cb}\frac{4G_F}{\sqrt{2}}
\left[ 
(\bar{c}_L \gamma_{\mu} b_L )+
\xi (\bar{c}_R \gamma_{\mu} b_R  ) 
\right] \left[ (\bar l_L \gamma^{\mu} \nu_{L})+
\xi^{\prime} (\bar l_R \gamma^{\mu}\nu_R  ) 
\right],
\end{equation}
with $\xi=g_R^q /g_L $, $\xi^{prime}=g_R^l /g_L $.

We write the expression for 
$ \Gamma( B_c^{-} \rightarrow l^- {\bar \nu_{l}}) $
generated by the Lagrangian in eq.(10)
\begin{equation}
\Gamma( B_c^{-} \rightarrow l^- {\bar \nu_{l}})=
\frac{G_F^2}{8\pi}|V_{cb}|^2 M_{B_c} f_{B_c}^2 m_l^2 \left( 1-
\frac{ m_l^2}{ M_{B_c}^2}\right)^2 
\times
\left( 1-\xi \right)^2 \left( 1+\xi^{\prime 2}  \right).
\end{equation}
It is also seen  that a small admixture of  $V+A$ current  modified
the SM prediction by a lepton mass independent factor
$\left( 1-\xi \right)^2 \left( 1+\xi^{\prime 2} \right)$,
where the factor $\left( 1-\xi \right)^2$ stems from
$\left[ (\bar c_L \gamma_{\mu} b_L )+\xi(\bar c_R \gamma_{\mu} b_R )\right]$
and $\left( 1+\xi^{\prime2} \right)$ from
$\left[(\bar l_L \gamma_{\mu}\nu_{L})+\xi^{\prime2}(\bar l_R \gamma_{\mu}\nu_R )\right]$.
It is obvious that a small $(V+A)$ admixture of lepton current would
induce a negligible small modification to SM predictions. This result can
be applied to $B^{\pm}$ and $ D^{\pm}$ leptonic decays. In what fellows,
we will neglect $\xi^{\prime2}$.

In ref.[16], it is shown that a small admixture of
$\left[ (\bar c_L \gamma_{\mu} b_L )+\xi(\bar c_R \gamma_{\mu} b_R )\right]$
modifies the rate of $B$ semileptonic decay by a factor 
\begin{equation}
r= \left( 1-0.74\frac{\xi}{1+\xi^2}\right) \approx \left( 1- 0.74\xi \right).
\end{equation}
However, in the  present case,
$\left[ (\bar c_L \gamma_{\mu} b_L )+\xi(\bar c_R \gamma_{\mu} b_R )\right]$
modifies the width of $B_c$ leptonic decay by
$\left( 1-\xi \right)^2 \approx \left( 1-2\xi \right) $.
For $\xi =0.14$, the width of B semileptonic decay is reduced by $10\%$[16],
however,  $\Gamma\left( B_c \rightarrow l\nu \right) $ will be reduced by about $28\%$.
The numerical results are 
displayed in Fig.3 as an illustration. 

It may be difficult to separate between the leptonic decays of $B_u$ and
$B_c$ at LHC just as at  LEP[17]. However, to our opinion, it might be possible
to separate between the processes
$b^{\star}\rightarrow B_c X_c \rightarrow\tau\bar\nu X_{c (s)}$ and 
$b^{\star}\rightarrow B_u X \rightarrow\tau\bar\nu X$ 
since the former one could be tagged by heavy flavor charm or strange
from $c\rightarrow s$. Anyway, it is a very complicate problem both for
theory and experiment, further discussion would be beyond the scope of 
this letter.

In summary, in contrast to other heavy mesons,
$B_c$ leptonic decay could be well studied and predicted  theoretically 
based on HQET and NRQCD. 
The experimental prospect at LHC is  potentially promising and encouraged.
We have  enough reasons to expect  that purely 
leptonic $B_c$ decays will offer an unique 
probe of the  SM and its extensions.

\par 
\bigskip
\noindent
{\large\bf Acknowledgment}

\noindent
This work is supported 
 in part by the National Natural Science Foundation and the 
Grant of State Commission of Science and Technology of China.

\bigskip

{\small

\newpage
\begin{center}
{\large Figure Captions}
\end{center}

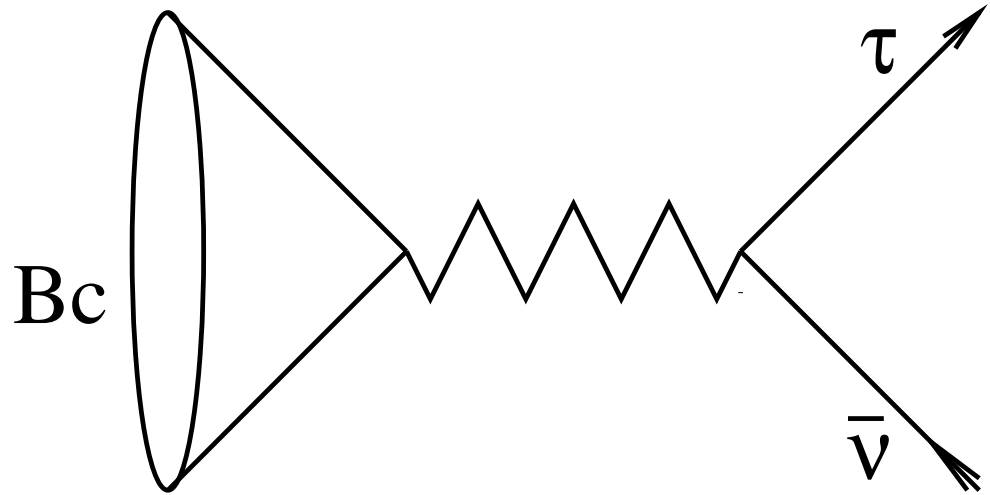
Fig.1. Diagram for $B_c$ annihilating into lepton pairs $via$ a virtual
$W^-$.  The shaded oval represents the wave function of $B_c$.
\vskip 2cm

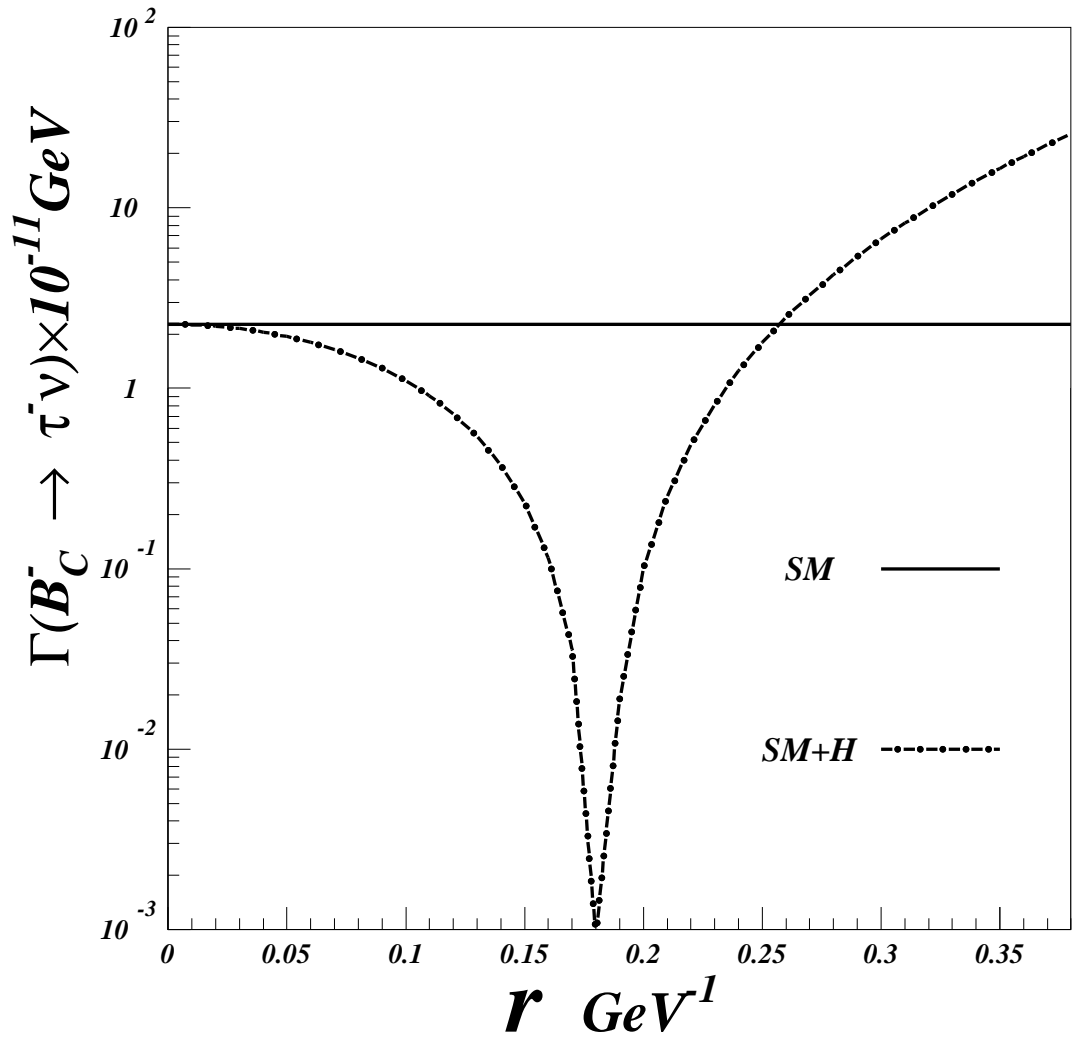
Fig.2. $\Gamma(B_c^- \rightarrow \tau\nu)$ as a function of 
$r=tan\beta/M_{H^\pm}$. 
The solid line is the SM prediction and the  dotted-dash line is the 
results including charged Higgs effects.
\vskip 2cm

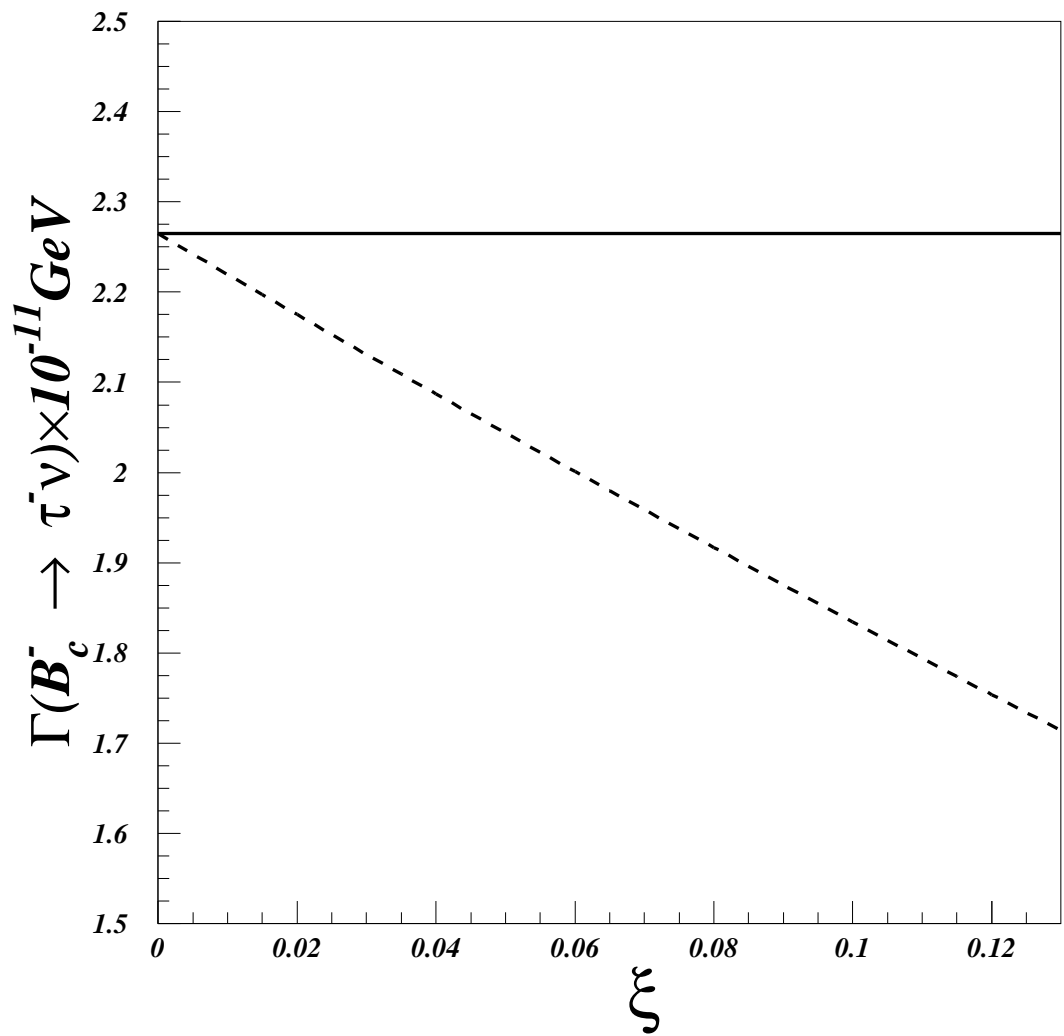
Fig.3. $\Gamma(B_c^- \rightarrow \tau\nu)$ as a function of $\xi =g_R /g_L$.
The solid line is the SM prediction and the dash line is the 
results of a small admixture of (V+A) quark current.

\end{document}